\documentclass{aa}  

\usepackage{graphicx}
\usepackage{txfonts}
\usepackage{subfig}
\usepackage{hyperref}
\usepackage{color}
\usepackage{comment}
\usepackage{xcolor}
\hypersetup{
    colorlinks=true,        
    linkcolor=blue,         
    filecolor=magenta,      
    urlcolor=cyan,          
    citecolor=black,         
    pdftitle={Overleaf Example},
    pdfpagemode=FullScreen,
}

\usepackage{xspace}
\newcommand{\HST}{HST\xspace}

\begin{document} 

\title{HOLISMOKES XV. Search for strong gravitational lenses combining ground-based and space-based imaging}
\titlerunning{Lens search combining two resolution images.}

   \author{A.~Melo
          \inst{1,2}
          \and
          R.~Cañameras\inst{3,1,2}
          \and
          S.~Schuldt\inst{4,5}
          \and
          S.~H.~Suyu\inst{2,1}
          \and 
          Irham T. Andika\inst{2,1}
          \and
          S. Bag\inst{2,1}
          \and 
          S. Taubenberger\inst{2,1}
          }

   \institute{Max-Planck-Institut für Astrophysik, Karl-Schwarzschild-Str. 1, 85748 Garching, Germany\\ 
   \email{amelo@mpa-garching.mpg.de}
   \and Technical University of Munich, TUM School of Natural Sciences, Physics Department, James-Franck-Straße 1, 85748 Garching, Germany
   \and Aix Marseille Univ, CNRS, CNES, LAM, Marseille, France
   \and Dipartimento di Fisica, Universita degli Studi di Milano, via Celoria 16, I-20133 Milano, Italy
   \and INAF - IASF Milano, via A. Corti 12, I-20133 Milano, Italy
   }

\date{Received -- --, ----; accepted -- --, -----}

 
\abstract{In the past, researchers have mostly relied on single-resolution images from individual telescopes to detect gravitational lenses. We propose a search for galaxy-scale lenses that, for the first time, combines high-resolution single-band images (in our case the Hubble Space Telescope, \HST) with lower-resolution multi-band images (in our case Legacy survey, LS) using machine learning. This methodology aims to simulate the operational strategies that will be employed by future missions, such as combining the images of Euclid and the Rubin Observatory's Legacy Survey of Space and Time (LSST). To compensate for the scarcity of lensed galaxy images for network training, we have generated mock lenses by superimposing arc features onto \HST images, saved the lens parameters, and replicated the lens system in the LS images. We test four architectures based on ResNet-18: (1) using single-band \HST images, (2) using three bands of LS images, (3) stacking these images after interpolating the LS images to HST pixel scale for simultaneous processing, and (4) merging a ResNet branch of \HST with a ResNet branch of LS before the fully connected layer.
We compare these architecture performances by creating Receiver Operating Characteristic (ROC) curves for each model and comparing their output scores. At a false-positive rate of $10^{-4}$, the true-positive rate is $\sim$0.41, $\sim$0.45, $\sim$0.51 and $\sim$0.55, for \HST, LS, stacked images and merged branches, respectively. Our results demonstrate that models integrating images from both the \HST and LS significantly enhance the detection of galaxy-scale lenses compared to models relying on data from a single instrument. These results show the potential benefits of using both Euclid and LSST images, as wide-field imaging surveys are expected to discover approximately 100,000 lenses.}


\maketitle

\section{Introduction}

Gravitational lensing occurs when a massive object, such as a galaxy or cluster of galaxies, bends and magnifies the light from a more distant object, such as a quasar or another galaxy. This effect can produce multiple images, arcs, or even rings of the background object in the case of strong lensing, providing a powerful tool for studying the distribution of dark matter and understanding the nature of dark energy \citep{1992Schneider}. Moreover, lensed time-variable objects like quasars or supernovae (SNe) can be used to measure the Hubble constant ($H_0$) and cosmological parameters, as proposed by \citet{1964Refsdal}. Our HOLISMOKES program, which stands for Highly Optimized Lensing Investigations of Supernovae, Microlensing Objects, and Kinematics of Ellipticals and Spirals \citep{2020Suyu}, aims to discover lensed SNe, determine the expansion rate of the Universe and study the progenitor systems of the SNe. The latter is possible by capturing the SN image at very early times \citep{2022Chen}. While only a few lensed SNe have been identified so far \citep{2015Kelly,2017GoobariPTF,2022KellyMACS2129,2022GoobarZwicky,2023PollettaSNH0pe,2024FryeSNH0pe,2021RodneySNRequiem,2022Chen,2024Pierel}, the number is expected to increase significantly with the wide-field imaging survey by the Rubin Observatory Legacy Survey of Space and Time (LSST, \citealt{2008Ivezic}) and the Euclid space telescope \citep{2011Laureijs}. Forecasts for these upcoming surveys suggest a substantial increase by a factor of several hundreds in the discovery of lensed supernovae \citep{2010OguriMarshall,2019Wojtak,2019Goldstein,2024Arendse,2024Bag,2024SainzdeMurieta}.

Over the past few decades, significant efforts have been made to detect and study gravitational lenses. Deep, wide-scale surveys have played a crucial role in finding new gravitational lenses, such as the Sloan Digital Sky Survey \citep[SDSS,][]{2006Bolton,2006Oguri,2012Brownstein}, the Cosmic Lens All-Sky Survey (CLASS; e.g., \citealt{2003Myers}), and the Survey of Gravitationally-lensed Objects in HSC Imaging (SuGOHI, \citealt{2018Sonnenfeld, 2020Chan, 2020Jaelani, 2022Wong,2023Jaelani,2024Chan}). The development of automated algorithms and machine learning techniques has revolutionized the field, allowing for processing vast amounts of astronomical data and identifying lens candidates with great efficiency and accuracy (e.g., \citealt{2018Lanusse, 2019Jacobs}). In particular, Convolutional Neural Networks (CNNs) have become increasingly important in astronomy due to their effectiveness in processing multi-wavelength imaging data (e.g., \citealt{2019Jacobs,2019Metcalf,2019Petrillo, 2020CañamerasHII, 2022Rojas,2024Schuldt}).

The Hubble Space Telescope (\HST) has been crucial in providing high-resolution images for detailed studies of known gravitational lenses and confirming new ones. Its capabilities have enabled researchers to discern intricate details in lens systems, resulting in improved constraints on lens models and the properties of the lensed sources \citep[see, e.g., the review by][]{2010Treu}. \citet{2008Faure} identified lens candidates in the COSMOS field \citep{2007Capak,2007Scoville}, while \citet{2018Pourrahmani} expanded on this by employing a CNN to search for lenses on single-\textit{HST}/ACS $i$-band observations of the COSMOS field. However, their resulting sample of strong-lens candidates suffered from high contamination, and the lack of color information made the visual inspection challenging. Single-band lens finding has limitations even at high resolutions, arguing for the inclusion of color information, even from lower-resolution ground-based images. Recent advancements have involved the use of multi-band imaging from ground-based surveys, such as the Dark Energy Survey (DES), the Legacy Survey (LS), PanSTARRS, and the Hyper Suprime-Cam (HSC) (e.g., \citealt{2017Diehl, 2020CañamerasHII, 2020Huang, 2021CañamerasHSVI, 2022Rojas}). These surveys have helped identify new lens candidates through color and morphological analysis. However, all the images are from one single instrument, and none have combined images of different resolutions and in different bands to search for strong lenses in an automated way.

Looking ahead, Euclid's wide-field space telescope is conducting an extensive survey, expected to cover approximately 14,000 square degrees of the sky \citep{2024MellierEuclid}. Being the first telescope to provide high-resolution images, even though just in a single optical band, for a large sky area, the large Euclid data set of galaxy images is suitable for gravitational lens searches \citep{2011Laureijs}. The searches of galaxy-scale lenses have already begun with the Early Release Observations data (\citealt{2024Acevedo-barroso}; Pearce-Casey et al., in preparation; and Chowdhary et al., in preparation). Furthermore, high-resolution images show great promise, particularly in the discovery of a wider range of lens configurations (e.g., \citealt{2022Garvin,2022Wilde}). Implementing advanced CNN architectures into the Euclid pipeline will be essential for efficiently identifying new lenses and maximizing the scientific output of the mission \citep{2019Petrillo}. Even though these high-resolution images are essential for resolving detailed lensing features, such as multiple images and faint arcs, multiband imaging will increase the efficiency and reduce false positives \citep{2019Metcalf}. 

Rubin LSST will complement Euclid's high-resolution imaging by providing deep, multiband observations across six optical bands: $u$, $g$, $r$, $i$, $z$, and $y$ \citep{2019Ivezic}. The LSST's ability to capture the sky repeatedly over its 10-year survey will create an unprecedented time-domain data set essential for identifying transient events and monitoring changes in known objects. By combining LSST's multiband data with Euclid's high-resolution images, the efficiency of gravitational lens searches should be significantly enhanced. This synergy between Euclid and LSST will create a powerful, complementary tool for the robust detection and confirmation of galaxy-scale lenses.

In this study, we present a CNN architecture based on ResNet-18 \citep{2016He}, trained on simulated and real lensing data, to improve the accuracy and efficiency of gravitational lens searches. While training solely on simulated data can lead to issues such as domain mismatch and overfitting to idealized conditions \citep{2018Lanusse,2018Schaefer}, we aim to overcome this by building a training set that incorporates real galaxy images. By combining high-resolution \HST data with multi-band LS images for the first time, our approach aims to improve detection capabilities, reduce false positives, and facilitate future applications in the Euclid and LSST missions.

This paper is organized as follows. In Sect.~\ref{section:Dataprep}, we introduce the construction of the ground truth data sets for training our CNNs, which includes simulating galaxy-scale lenses (for \HST and LS).  The network architectures and training are presented in Sect.~\ref{section:Networktraining}. The results of applying the networks are presented in Sect.~\ref{section:resultsanddiscussion}, along with a discussion about the architecture's performance, specially when combining images of different resolution. Finally, our main conclusions are in Sect.~\ref{section:Conclusions}.

\section{Data preparation}\label{section:Dataprep}

\begin{figure}[!htbp]
    \centering
    \includegraphics[width=0.5\textwidth]{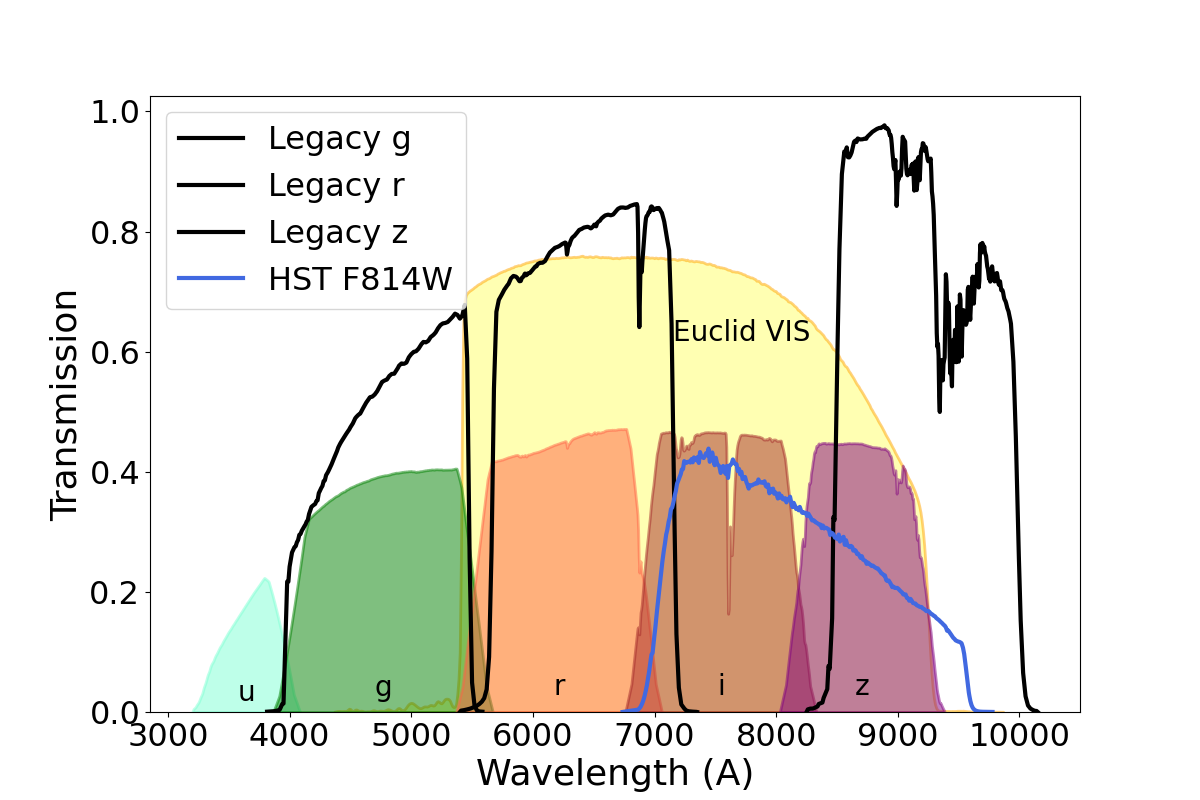}
    \caption{Some of the filters from LSST (in shaded color labeled with $u$, $g$, $r$, $i$, and $z$) and Euclid (shaded yellow). The Legacy Survey and \HST F814W filters are shown in black and blue lines, respectively. Image based on \citet{2019Capak}.}
    \label{fig:filters}.
\end{figure}

We simulated a sample of galaxy-scale lenses with two different resolutions, corresponding to a scenario with single-band space-based imaging and multiband ground-based imaging. Since Euclid provides high-resolution imaging primarily through its VIS channel (0.1\arcsec), which is most suitable for lens finding due to its superior resolution compared to the NISP channels (0.3\arcsec) \citep{2015Collett}, we will use the \HST filter F814W image that has a pixel scale of 0.05\arcsec. For low-resolution multi-band imaging, we use images from the DECam Legacy Survey (DECaLS) in the filters $g, r$, and $z$, which have a pixel scale of 0.263 arcseconds. These filters are selected to facilitate comparison with filters from the LSST and Euclid survey (see Fig.~\ref{fig:filters} for comparison), and are driven by the options currently available.  

Galaxy-scale lenses need to be simulated due to insufficient data for network training (deep learning typically requires $>$10,000 training data). The process involves creating realistic arcs around real galaxies from LS and \HST images, which act as lenses. It requires several steps: selecting galaxies that act as the lenses, choosing the source images, and performing a strong lensing simulation. To ensure consistency between high-resolution and low-resolution mock images, we use the same lens parameters and source positions, creating identical lens configurations across resolutions.

\begin{figure*}[!htb]
    \centering
\includegraphics[width=0.9\linewidth]{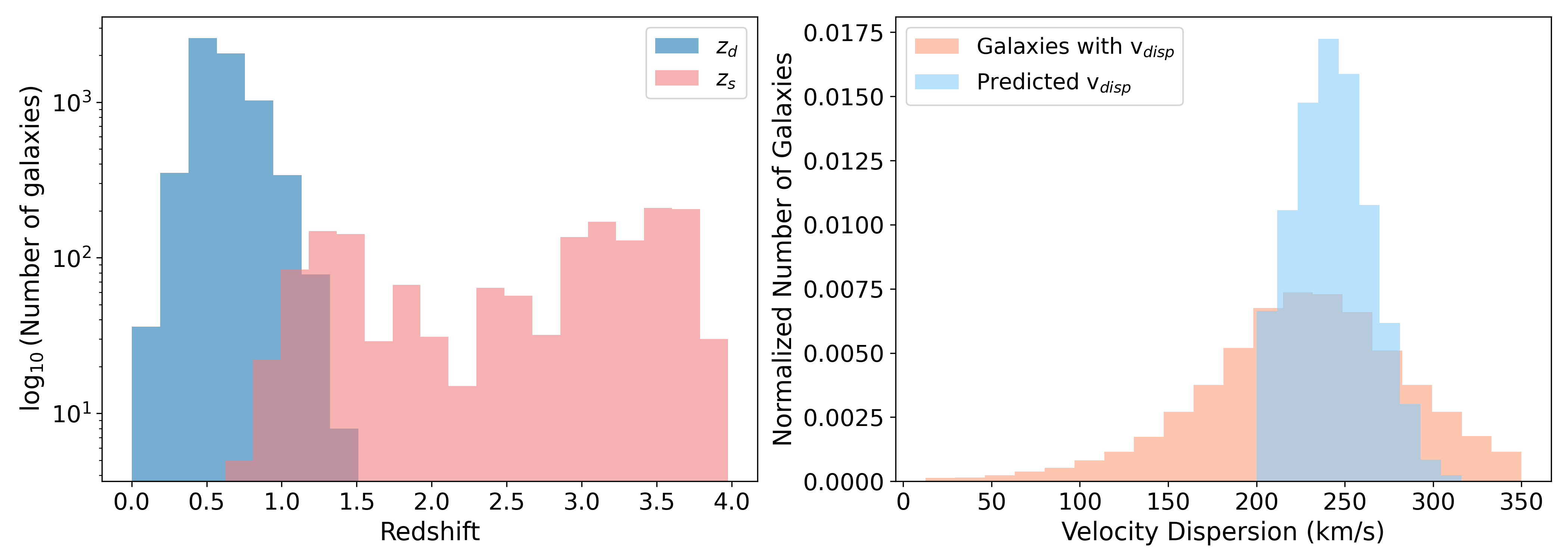}
    \caption{Galaxy parameters ($z_\text{d}$, $z_\text{s}$, and $v_{\mathrm{disp}}$) of the lenses and sources used for simulating galaxy-scale lenses. $(Left)$ Histogram of the number of galaxies as a function
    of the redshift of the source $z_\text{s}$ (red) and lens $z_\text{d}$ (blue). $(Right)$ Normalized $v_{\mathrm{disp}}$ distribution measured from SDSS used for training the KNN network (orange color), and the predicted normalized $v_{\mathrm{disp}}$ distribution for lens galaxies over the final sample of mocks (blue color).}
    \label{fig:galproperties}
\end{figure*}

First, we search for luminous red galaxies (LRGs) to use as the lenses. These galaxies are preferred due to their high lensing cross-section (e.g., \citealt{1984Turner}) and because their smooth and uniform light profiles make it easier to separate the background emission from the lensed sources. We identify LRGs in the Legacy Survey using the selection criteria from \citet{2023Zhou}, accessed through the Astro Data Lab's query interface\footnote{\url{https://datalab.noirlab.edu/}} (\citealt{2020Nikitta}). The color cuts in the selection criteria are based on the optical $grz$ photometry from the Data Release 9 (DR9) and the WISE (\citealt{2010Wright}) $W1$ band in the infrared. After obtaining the list of LRGs, we crossmatch their coordinates with the \HST Legacy Archive\footnote{\url{https://hla.stsci.edu/hlaview.html}}; in cases of successful matches, we keep objects whose corresponding \HST images have exposure times (total per object) of at least 100 seconds.
The relatively low time limit is required to obtain an \HST F814W image for at least $\sim$1/3 of the LS LRGs. With this criterion, we identify a total of approximately $\sim 10000$ LRG images in LS and \HST. We then download the $20\arcsec \times20\arcsec$ cutouts of the image.

\begin{figure}[!htb]
    \centering
\includegraphics[width=1.0\linewidth]{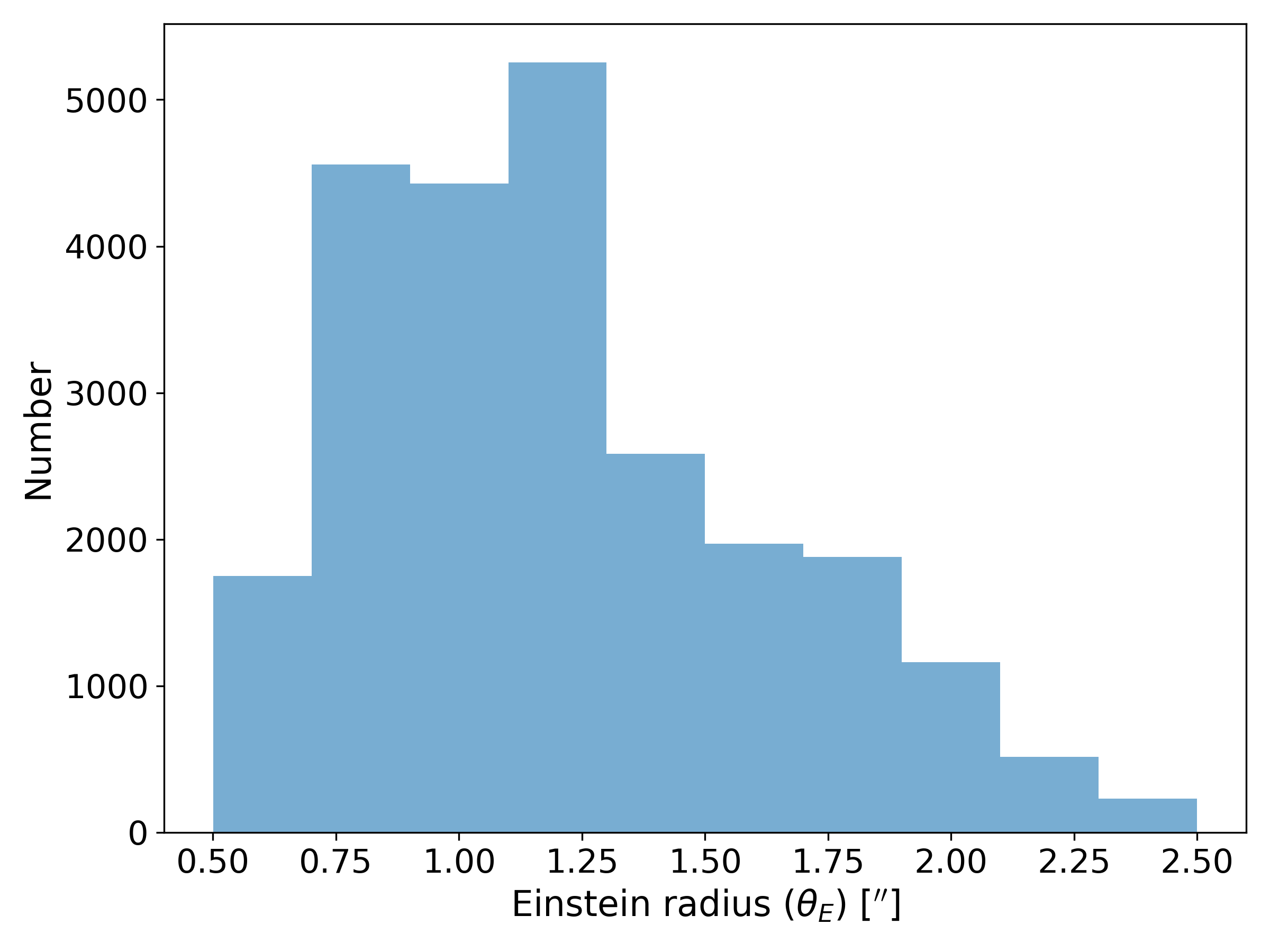}
    \caption{Distribution of the Einstein radii of the mock strong lens systems used to train our networks.}
\label{fig:ein_radius}
\end{figure}

For the background sources, we refer to the catalog created by \citet{2021Schuldt} of images of high-redshift galaxies from the Hubble Ultra Deep Field (HUDF, \citealt{2017Inami}). These galaxies are chosen due to their high spatial resolution (pixel size of 0.03\arcsec), high signal-to-noise ratio (SNR), availability of spectroscopic redshifts \citep{2006Beckwith,2017Inami}, and compatibility with the LS filters $g$, $r$, and $z$, corresponding to the \HST filters $F435W$ ($\lambda = 4343.4,\AA$), $F606W$ ($\lambda = 6000.8,\AA$), and $F850LP$ ($\lambda = 9194.4,\AA$), respectively. The spectroscopic redshifts of the selected HUDF galaxies falls in the range 0.1 $< z <$ 4.0. These images can be used for simulations for each filter in the LS images, with color corrections applied to account for differences in filter transmission curves and photometric zero points. Additionally, the HUDF $F775W$ filter is used for the simulation in \HST images, with color corrections to account for differences between filters.  

\begin{figure*}[!htb]
    \centering\includegraphics[width=\textwidth]{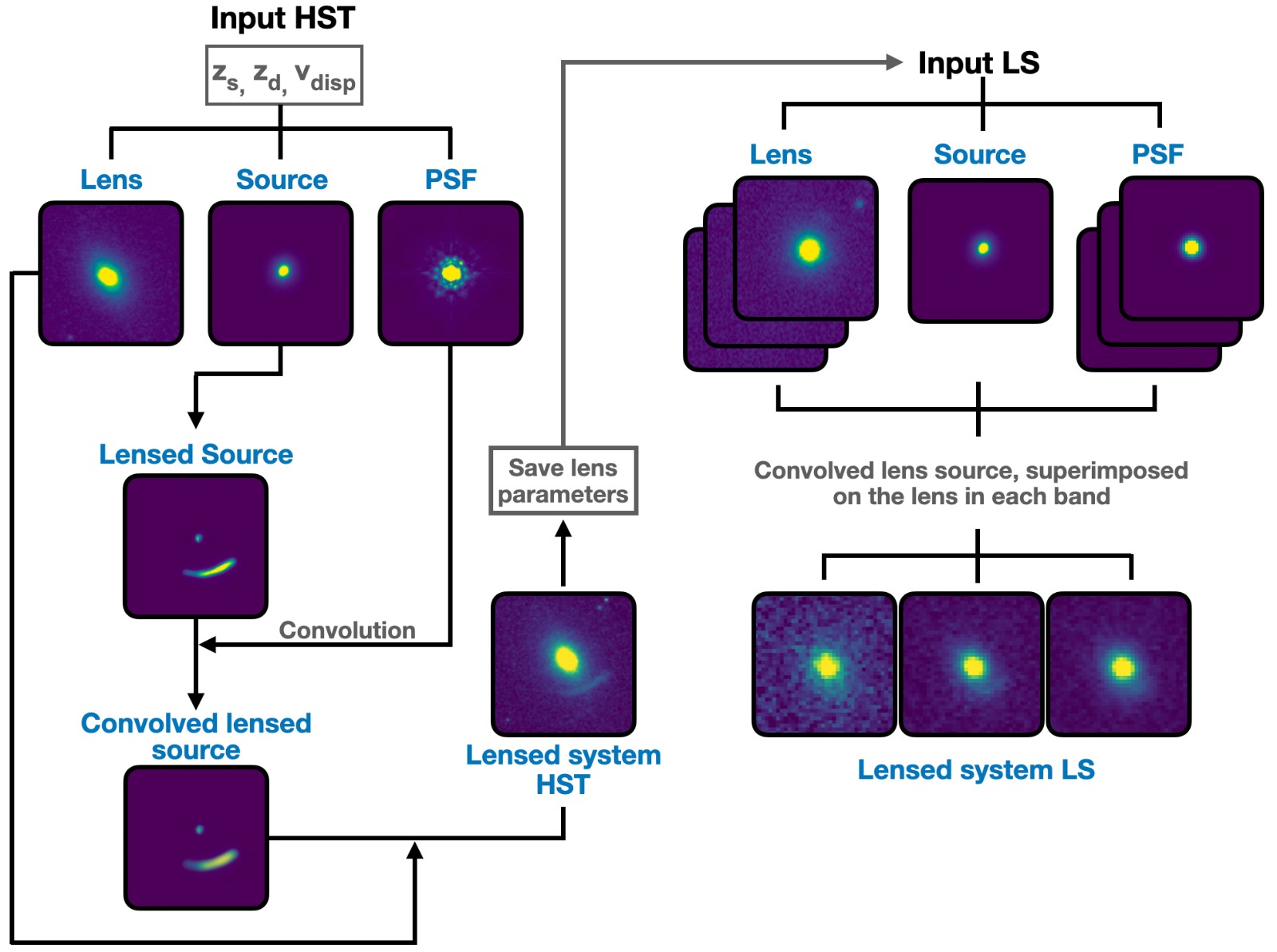}
    \caption{Simulation pipeline procedure used for simulating galaxy-scale lenses. See further details in Sec.~\ref{section:Dataprep}.}
    \label{fig:sim_schuldt}
\end{figure*}

We follow the procedure described in \citet{2021Schuldt} to simulate realistic galaxy-scale lenses by painting lens arcs first on the \HST images. As input parameters, we require the redshift of the source ($z_\text{s}$), the redshift of the lens ($z_\text{d}$), and the velocity dispersion of the deflector ($v_{\mathrm{disp}}$). Since not all LRGs have $v_{\mathrm{disp}}$ and redshift measured from the SDSS, we use the photometry and photometric redshift from LS to predict the velocity dispersion using a K-nearest neighbor (KNN) algorithm, following \citet{2022Rojas}. 
We train this algorithm with velocity dispersions using $\sim 580,000$ LRGs with redshifts from SDSS and photometry from LS from our original list of LS without corresponding \HST image. To keep the LRGs with reliable $v_{\mathrm{disp}}$, we remove LRGs with $v_{\mathrm{disp}}$ $\leqslant 350$ km~s$^{-1}$ and $v_{\mathrm{disp\_error}}$ $\leqslant 50$ km~s$^{-1}$ before training. As a result, we obtain a root-mean-square (rms) scatter in the prediction of $\sigma = 65~\text{km s$^{-1}$}$. The distribution of the $z_\text{d}$, $z_\text{s}$ and $v_{\mathrm{disp}}$ from the input lens and source catalogs are shown in Fig.~\ref{fig:galproperties}.

To create galaxy-scale lens simulations, we adopt a lens mass distribution with a Singular Isothermal Ellipsoid \citep[SIE;][]{KassiolaKovner93} profile using the input parameters: $z_\text{d}$, $z_\text{s}$, and $v_{\mathrm{disp}}$. The lens centroid, axis ratio, and position angle are derived from the first and second brightness moments in the \HST F814W images. The sources are randomly positioned in the source plane, and only simulations with an Einstein radius ($\theta_\text{E}$) between $0.5\arcsec$ and $2.5\arcsec$ are considered (see Fig.~\ref{fig:ein_radius}). The choice of the range in $\theta_\text{E}$ takes advantage of the high resolution provided by \HST to study whether the architecture can effectively identify lensed systems with small $\theta_\text{E}$, which were previously challenging to differentiate from non-lenses when using single-resolution images. Subsequently, the sources are lensed in the image plane using the GLEE software \citep{2010Suyu,2012Suyu}. Finally, for the \HST mock simulation, the simulated lensed source image is convolved with the \HST Point Spread Function (PSF) model created using the Tiny Tim package \citep{1995Krist}, then rescaled using the \HST zero-points, and added to the original LRG image to produce the final mock simulation. We refer to \citet{2021Schuldt} for additional details on the procedure. To simulate the same lens for ground-based multiband imaging of LS, we store both the lens parameters and the fixed source position from the \HST images, and use them to create the same lens configuration using the GLEE software for the LS $g$, $r$, and $z$ images. We make sure to resample and rescale the \HST source images using the LS PSF (computed using PSFEx and available through the Legacy Survey Viewer\footnote{\url{https://www.legacysurvey.org/viewer}}) The simulation process is summarized in Fig.~\ref{fig:sim_schuldt}.

\begin{figure*}
    \centering
    \includegraphics[width=0.75\textwidth]{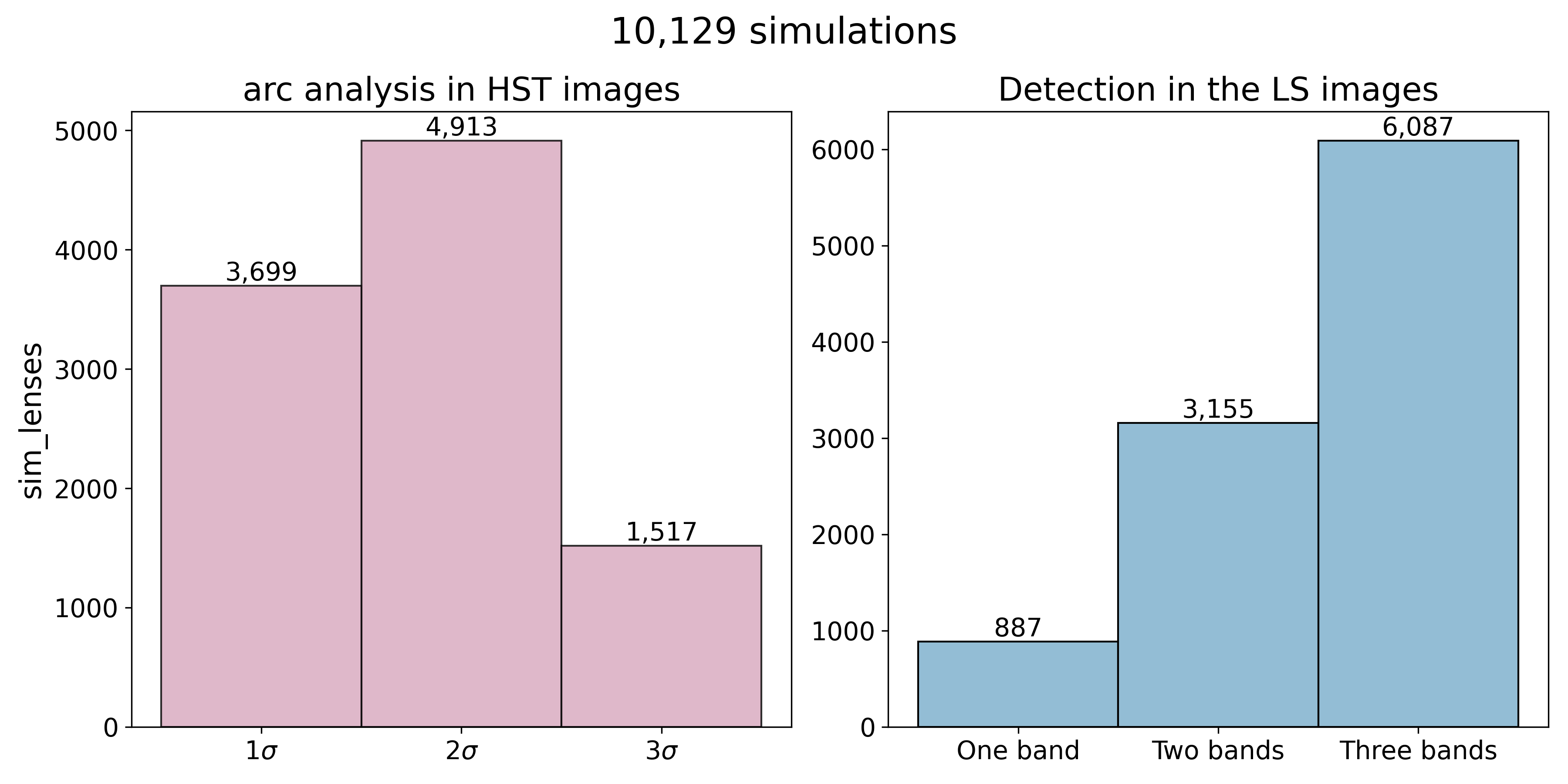}
    \caption{Properties of the 10,129 \HST and LS lens simulations: $(Left)$ number of mock lenses whose integrated arc brightness is 1$\sigma$, 2$\sigma$ and 3$\sigma$ above the sky background in \HST images,
    and $(Right)$ number of mock lenses whose arcs are detected in LS images above 2$\sigma$. There are $\sim$6,000 mocks whose arcs are detected at $>2\sigma$ in \HST and in all three LS bands.}
    \label{fig:arc_analisis}
\end{figure*}

We initially obtain a total of 10,129 simulated lenses. However, upon comparing the simulated simulated HST with the simulated LS images, we observe that not all arcs are visible in the LS images when using the same lens parameters. This outcome is expected due to the use of images with different resolutions and colors. To ensure the consistency in our sample, we select simulations where the arcs are visible in all three LS bands and are at least 2$\sigma$ above the sky background in the \HST images (see Fig.~\ref{fig:arc_analisis} for further clarification of the whole sample). This leaves us with a total of $\sim$6,000 simulations.

Since this number is insufficient for training a network, we employ image rotation to the original LRG images (90, 180, and 270 degrees) and simulate new galaxy-scale lenses using different source images in each case. This creates additional positive examples and effectively increases the data set size to approximately $\sim$24,000 positive images. This process not only increases the size of the data set but also enhances its variability, helping the model to become more robust to variations in orientation and configuration, which improves its generalization capabilities when applied to real astronomical data. The Einstein radius of the entire $\sim$24,000 mock systems with HST and LS images is shown in Fig.~\ref{fig:ein_radius}; the overall distribution peaks at $\theta_E$ between $0.8$ and $1.2$\arcsec.

We use a diverse set of non-lens contaminants for the negative examples to help the networks learn the diversity of non-lens galaxies, with a particular focus on morphological types that resemble strong lenses (see \citealt{2022Rojas,2023Cañameras}). This includes various categories such as LRGs, ring galaxies, spiral galaxies, edge-on galaxies, groups, and mergers. The LRGs are obtained from the same parent sample that provides the lenses. Coordinates for the spiral and elliptical galaxies are obtained from the Galaxy Zoo\footnote{\url{https://data.galaxyzoo.org/}} and then used to download the images in \HST and LS. In addition, galaxy classifications from the Galaxy Zoo: Hubble (GZH) project \citep{2017Willett} and Galaxy Zoo DECaLS \citep{2022Walmsley} are used to identify and download rings and mergers. All these negatives are then split into the training, validation, and test sets.

\begin{figure*}
    \centering
    \includegraphics[width=1.0\linewidth]{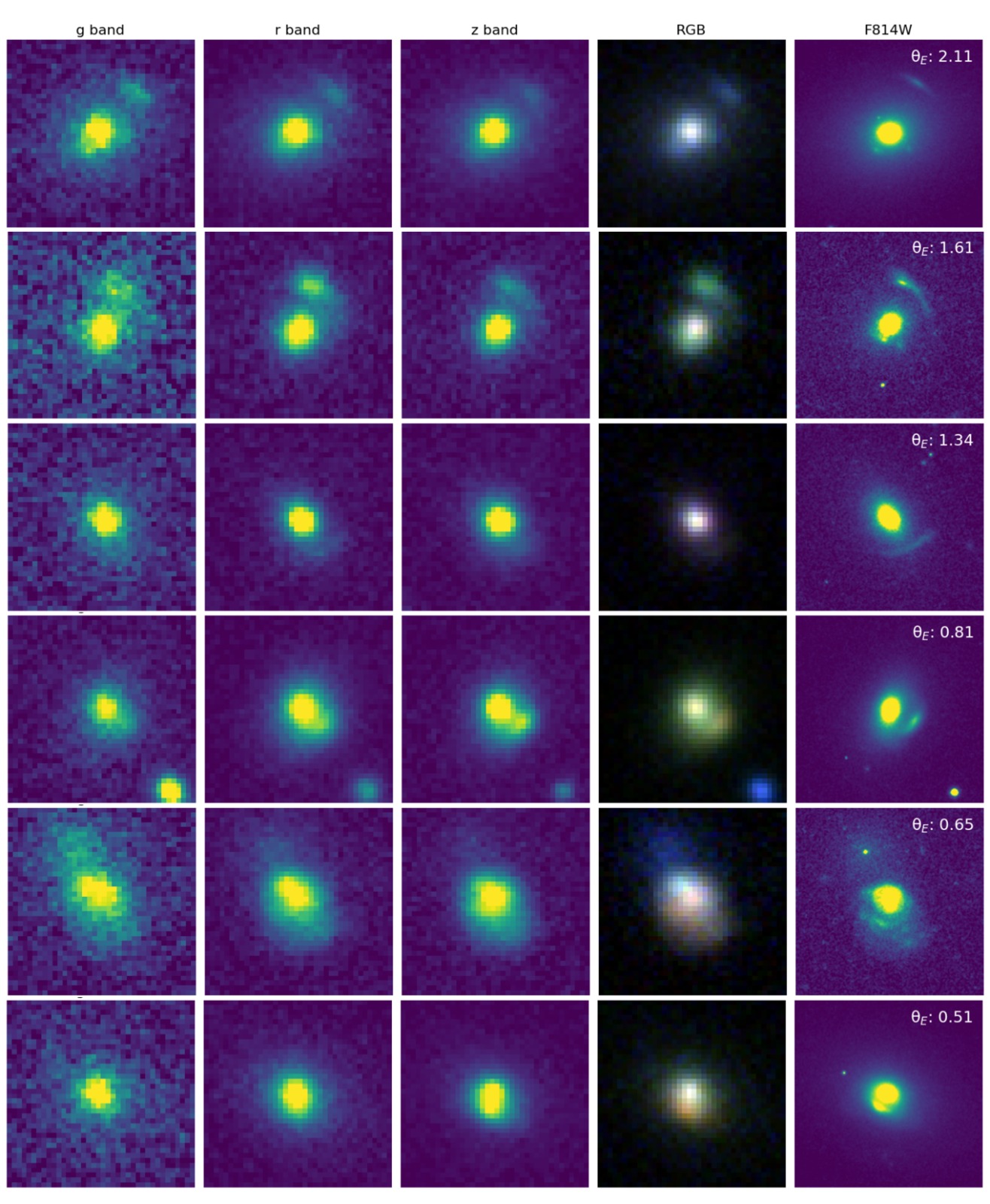}
    \caption{Positive lens simulations of LS and \HST for different Einstein radii (in descending order). From left to right: $g$, $r$, and $z$ band images, plus the RGB image from LS. The last panel on the right shows the image in the F814W band from \HST. Cutouts have a size of 10\arcsec$\times$10\arcsec.}
    \label{fig:positives}
\end{figure*}

Finally, we split the positive data sample into 56\% training, 14\% validation, and 30\% test sets (see Table~\ref{tab:hst_ls_dist}). For the test set, we include $\sim$8,000 positive examples, which consist of simulated galaxy-scale lenses, along with 364 real lens candidates obtained from the literature \citep{2008Bolton,2009Auger,2014Pawase}
that have both \HST and LS images. Since we need a larger sample of negative examples for the test set in order to obtain robust estimates of low false-positive rates (FPRs), especially in \HST, we also use images of objects observed in the field-of-view (FOV) of the LRGs acting as the lenses. These contaminants are initially identified with a flux threshold above the background. Subsequently, the image is cropped to 10\arcsec$\times$10\arcsec ~ in both \HST and LS. Later, we filtered the images to ensure that there were no zero values in the field. While we acknowledge that using LRGs from the same simulations as negative examples and extracting cutouts from the same FOV is not ideal—leading to potential repetitions in training and test sets—this approach is driven by the limited data available to us. Despite these limitations, the negatives provide a diverse and representative sample for training. The $\sim$120,000 in the test set are a combination from both \HST and LS data. Examples of positive images can be seen in Fig.~\ref{fig:positives}. The final image used for training is 10\arcsec$\times$10\arcsec, resulting in 200$\times$200 pixels for \HST images and 38$\times$38 pixels for LS images.

\begin{table}[h]
\centering
\begin{tabular}{ |l|c|c|c| }
 \hline
 \HST/Legacy Survey & Training & Validation & Test set\\ 
 \hline
Positive & $\sim$14,300 & $\sim$3,600 & $\sim$8,000  \\  
Negative & $\sim$20,000 & $\sim$4,300 & $\sim$120,000 \\ 
 \hline
\end{tabular}
\caption{Distribution of samples in the \HST/Legacy Survey data set for training, validation, and test sets.}
\label{tab:hst_ls_dist}
\end{table}

\begin{figure*}[!htb]
\centering
    \begin{minipage}[b]{0.41\linewidth} 
        \centering
        \subfloat[Legacy Survey architecture.]
        {
            \label{a}
            \includegraphics[width=0.95\linewidth]{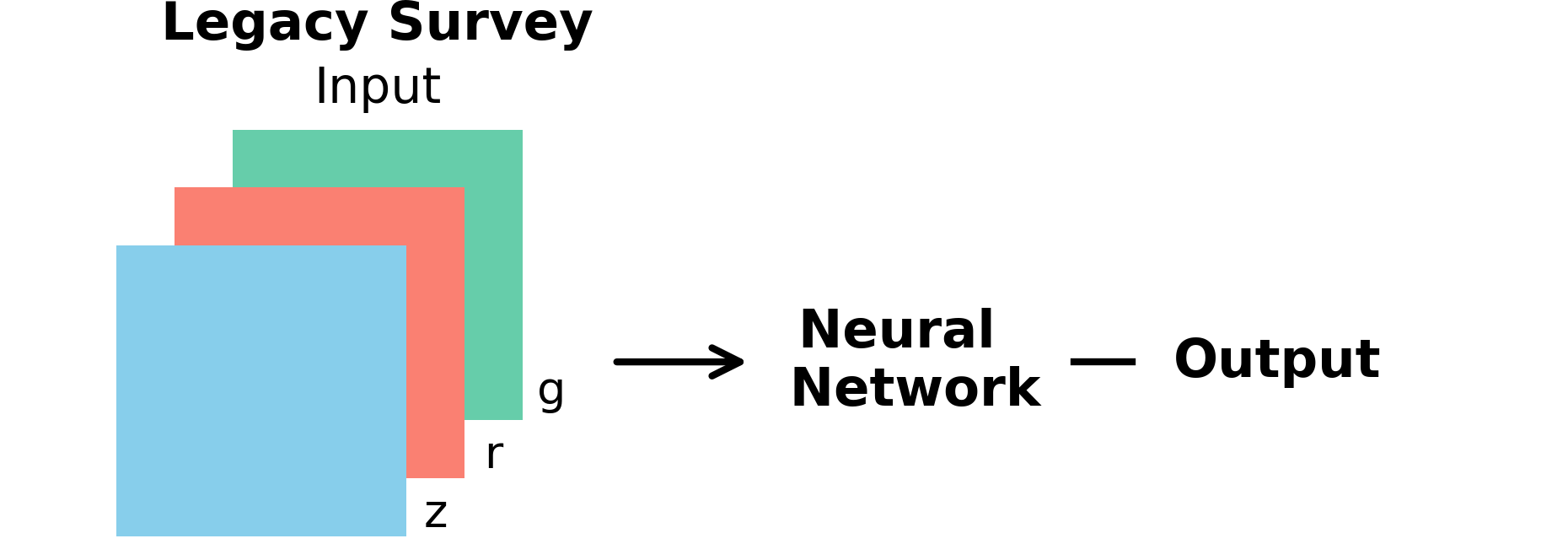}
        }\\  
        \subfloat[HST architecture.]
        {
            \label{b}
            \includegraphics[width=0.95\linewidth]{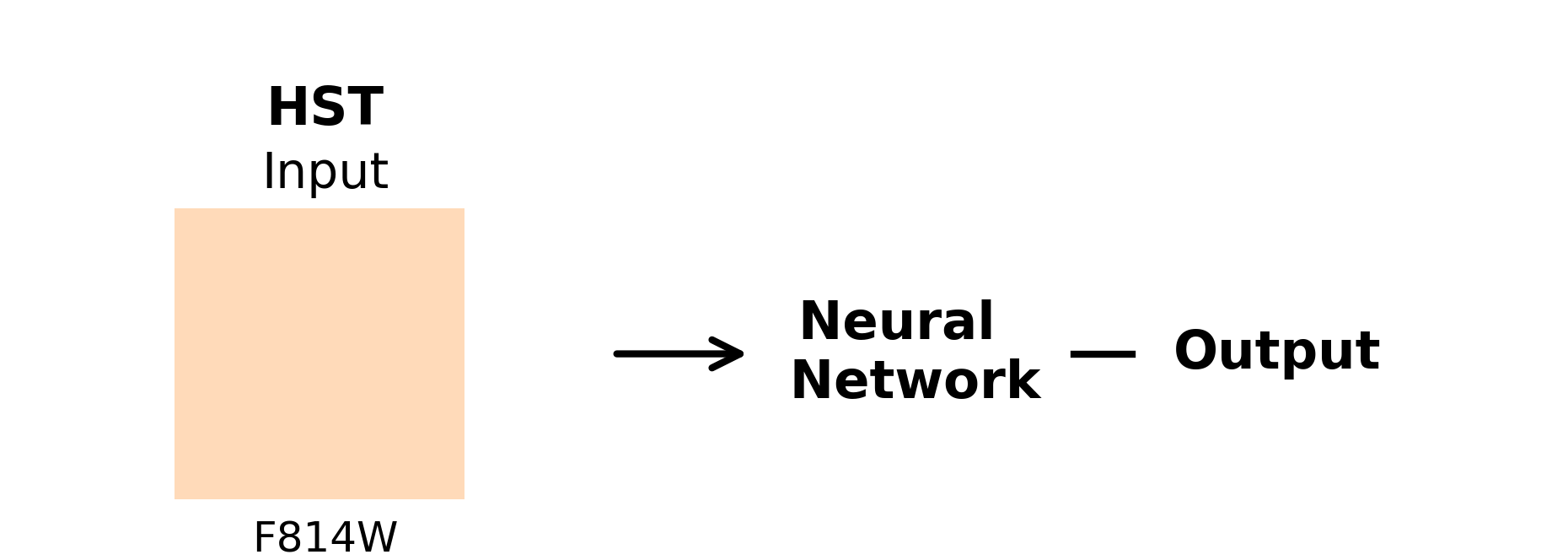}
        }
    \end{minipage}
    \hfill
    \begin{minipage}[b]{0.45\linewidth} 
        \centering
        \subfloat[Stacked-images architecture.]
        {
            \label{c}
            \raisebox{10mm}{\hspace{-20mm}\includegraphics[width=1.2\linewidth]{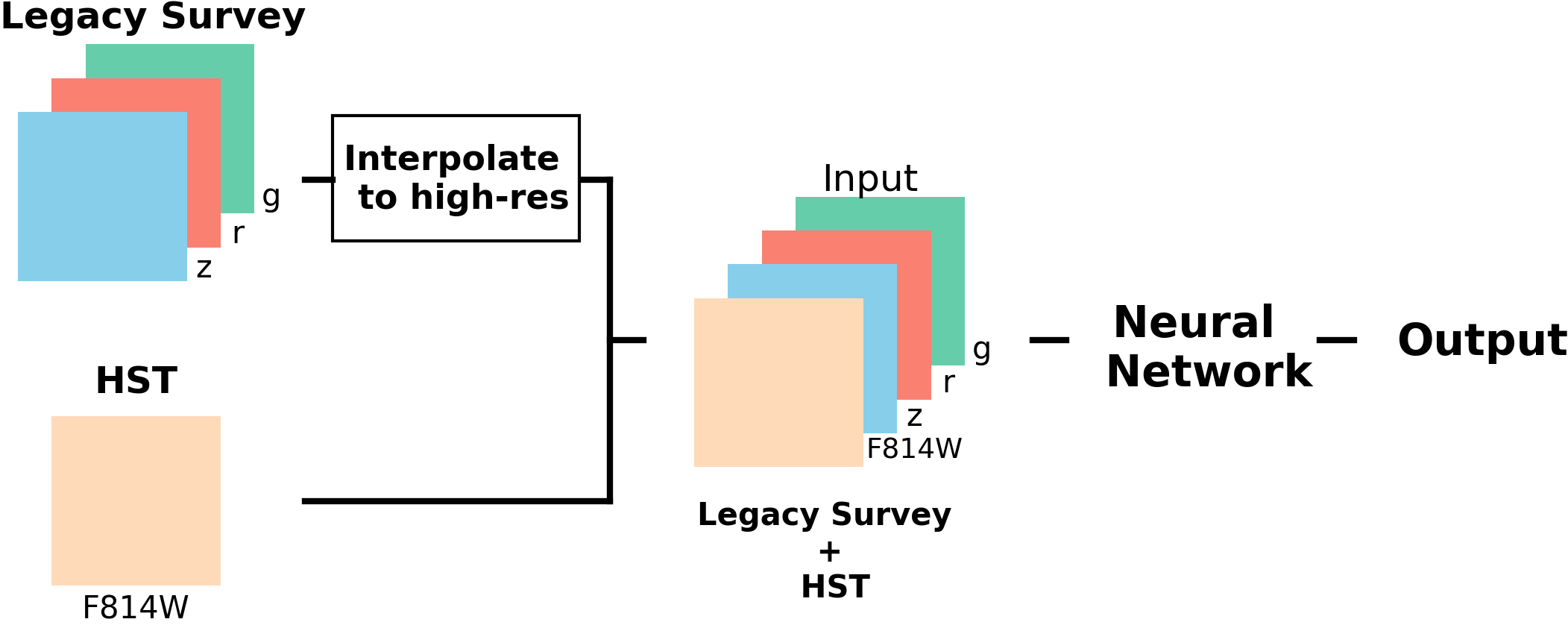}}
        }
    \end{minipage}
    
    \vspace{1em} 
    \begin{minipage}[b]{0.51\linewidth}
        \centering
        \subfloat[Merged-branches architecture.]
        {
            \label{d}
            \includegraphics[width=\linewidth]{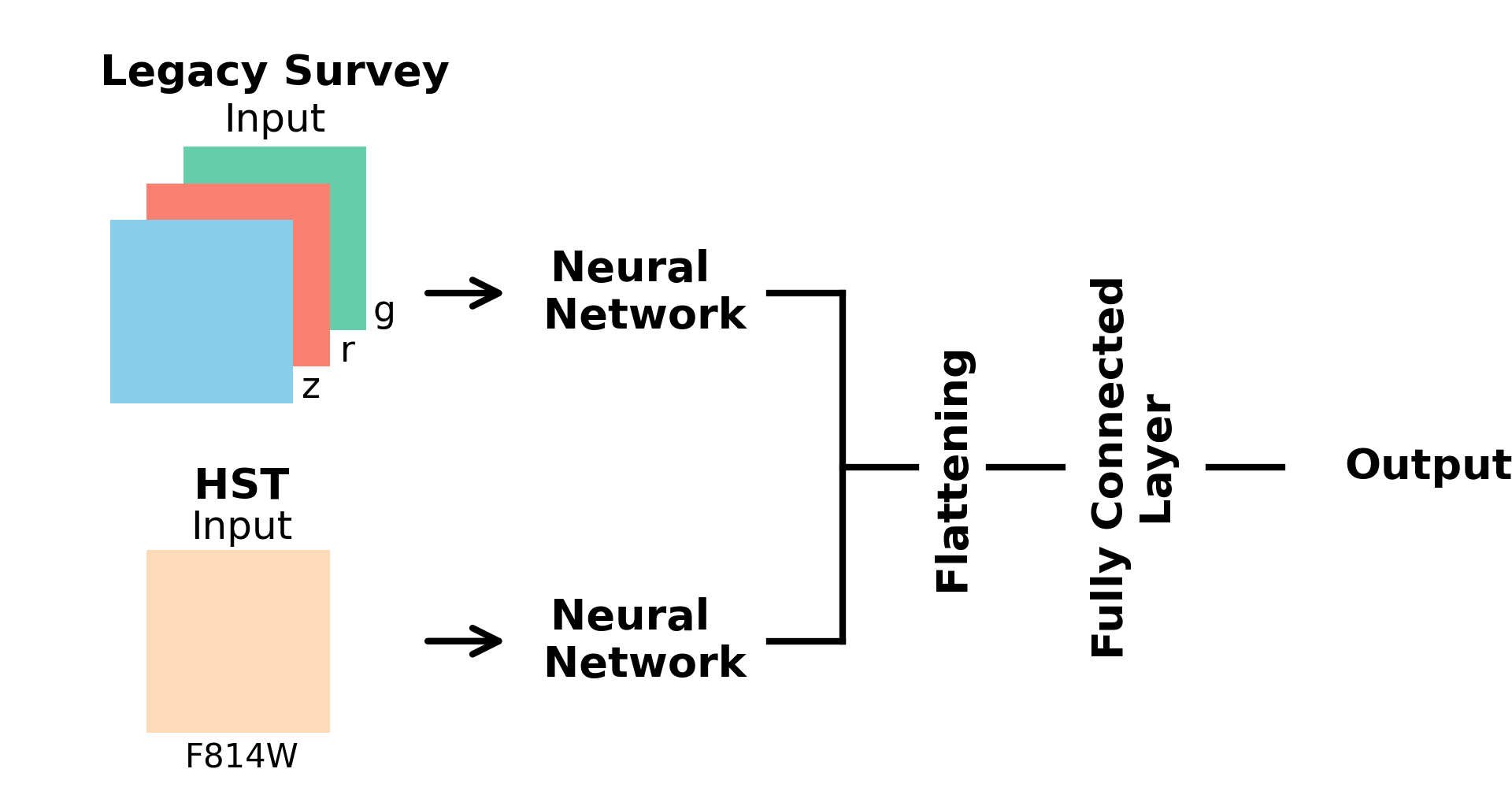}
        }
    \end{minipage}

    \caption{Visualized architectures for the four scenarios: (a) LS, (b) \HST architecture, (c) stacked-images architecture, and (d) merged-branches architecture.}
    \label{fig:architectures}
\end{figure*}

\section{Network training}\label{section:Networktraining}

We need our network to be both effective and simple to accommodate training for four scenarios (see Fig.~\ref{fig:architectures}):

\begin{enumerate}
    \item LS images (three images as input).
    \item \HST images (single image as input).
    \item Stacked images (four images as input)
    \item Merged branches (four images as input, but split into two branches)
\end{enumerate}

To accomplish the research objective, we use ResNet-18, a deep CNN architecture introduced by \citet{2016He}, in the four scenarios. This network has previously been used in lens search for instance by \citet{2021CañamerasHSVI}, who employed a network inspired by ResNet-18 to search for lenses in the HSC Survey, and also by \citet{2022Shu} (based on the work of \citealt{2018Lanusse}), who use it to search for high-redshift lenses in the HSC Survey.
The architectures are slightly modified to fit each specific scenario, as explained below.

The architecture begins with a 7x7 convolutional layer followed by max-pooling. It includes four groups of residual blocks, each containing two blocks; these groups sequentially use 64, 128, 256, and 512 filters. Each residual block has two 3x3 convolutional layers, with identity shortcuts connecting the input directly to the output. The network concludes with a global pooling layer and a fully connected layer for classification, followed by a sigmoid activation function that outputs a single value between 0 and 1, allowing efficient learning and improved performance in deep learning tasks. The model is optimized using the Stochastic Gradient Descent (SGD) optimizer, with categorical cross-entropy as the loss function, and is trained with a batch size of 128, ensuring stability and efficiency during training. 

While the tested network architectures using only HST images or only LS images (see Fig.~\ref{fig:architectures} images (a) and (b), respectively) follow a straightforward approach based on the modified ResNet-18 architecture from previous studies, the innovation in this work lies in how we combine these data sets. For the stacked-images architecture (see Fig.~\ref{fig:architectures}(c)), we use cubic interpolation to convert the low-resolution LS images to high-resolution images matching those from the \HST. This means that the final LS images have the same pixel resolution as the \HST images after interpolation, without altering the intrinsic content of the original images. By doing this, we can directly use the four images ($g, r, z$, and F814W) together as input to the network architecture. In the training process, the input data is balanced to ensure that LS and \HST contribute equally to the overall loss function. While the LS data set contains three bands ($g$, $r$, $z$) and \HST only one band, each band in LS is treated as an independent input, effectively increasing the sample size of LS. To avoid bias, the training process balances the number of samples used from each data set, ensuring equal representation during training.

In the merged-branches scenario (see Fig.~\ref{fig:architectures} (d)), both types of astrophysical images (\HST and LS) are processed through separate branches up to a point, but they ultimately contribute to a shared neural network. This includes a fully connected (FC) layer, meaning that while the initial layers handle the \HST and LS data independently, the final layers merge to form a single network that processes both data types together. We then combine the outputs from these networks to carry out the final classification. To elaborate, each set of images is run through its dedicated network (\HST images through a ResNet-18 model and LS images through another ResNet-18 model), with their feature maps flattened into vectors. We combine the flattened feature vectors to merge information from both networks into a new feature dimension. The input images of each object provide 512 features from the \HST ResNet-18 and another 512 features from the LS model, resulting in a final feature vector of 1024 dimensions. This extensive feature set presents challenges like increased computational complexity, higher memory requirements, and potential overfitting. To tackle these challenges, we use another fully connected layer to reduce the 1024-dimensional feature vector to a lower dimension. A ReLU activation function is then applied to the output of the reduction layer. Next, the reduced feature vector undergoes a fully connected (dense) layer to refine the feature representation further. Finally, the processed feature vector is input into the output layer, where the sigmoid activation function is applied to yield a probability score ranging from 0 to 1 for each input image. A score close to 1 indicates a high likelihood of the image showing a gravitational lens, whereas a score near 0 suggests the absence of strong lensing, or other galaxy types rather than a lens.

\begin{figure*}[!htb]
    \centering
    \includegraphics[width=0.48\linewidth]{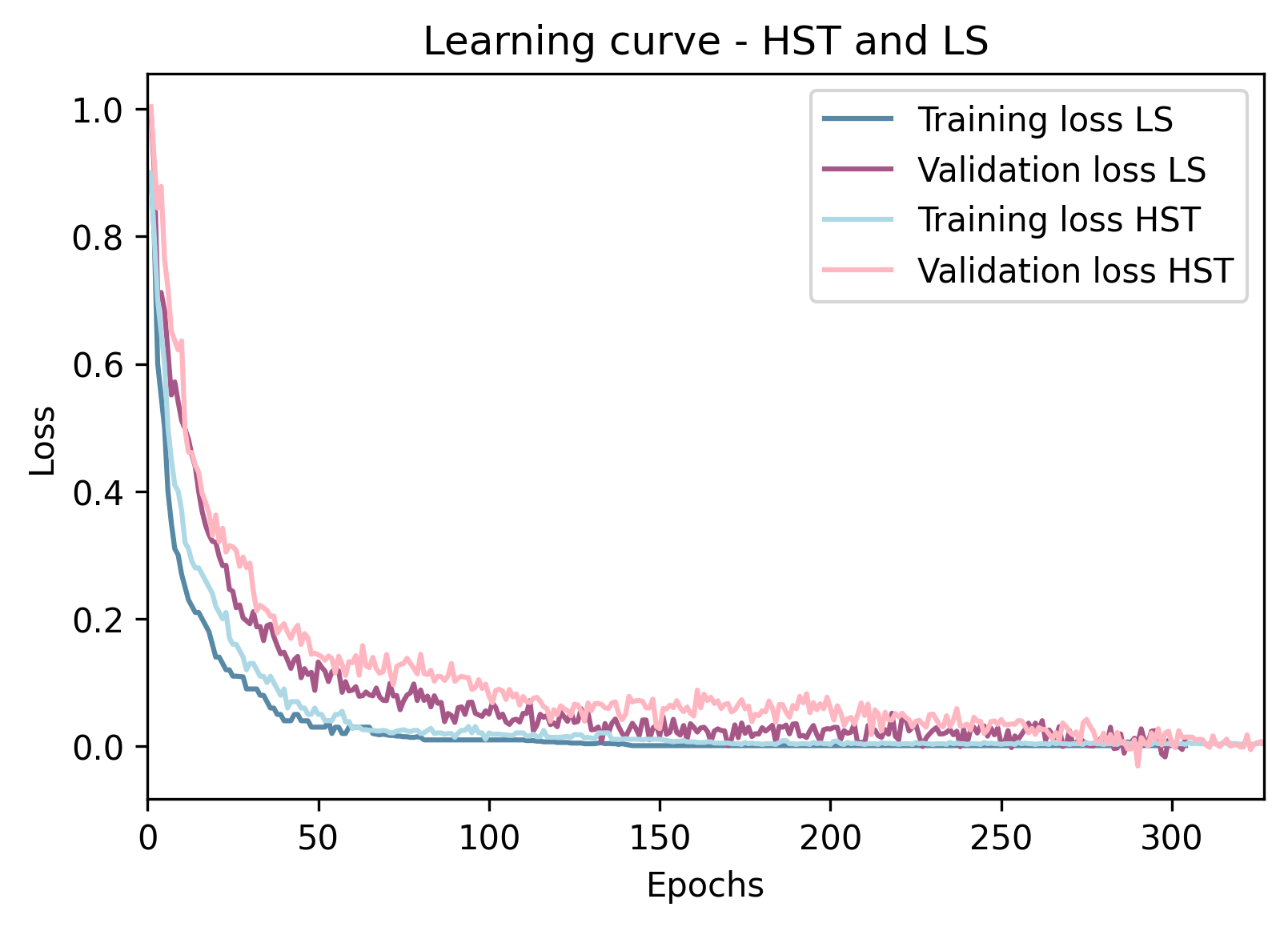}
    \includegraphics[width=0.48\linewidth]{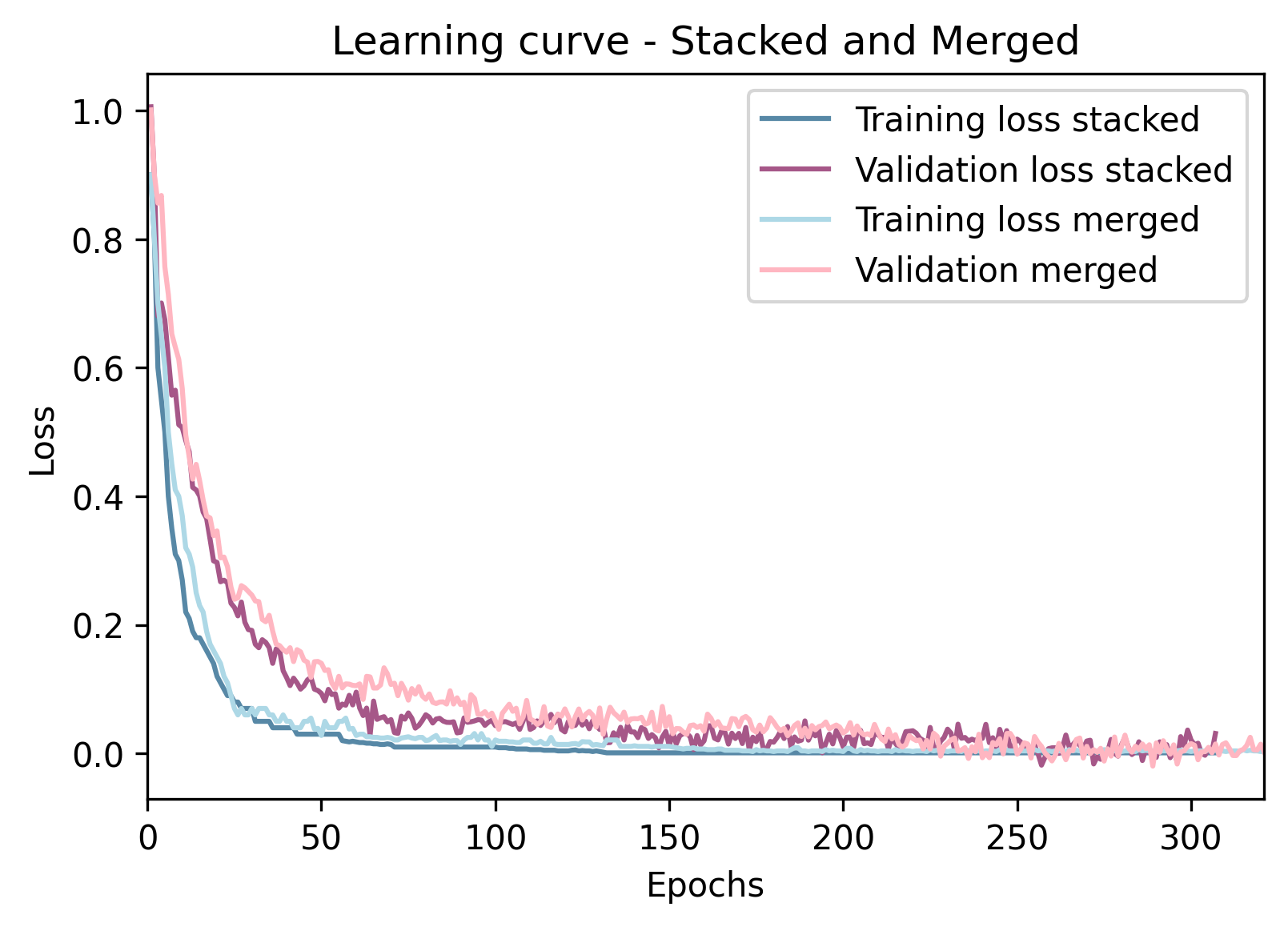}
    \caption{Loss curves for the training (blue and light blue colors) and validation (purple and pink colors) for the four scenarios. Left: \HST and LS alone. Right: stacked and merged branches.}
    \label{fig:loss_curves_1}
\end{figure*}

Early stopping was applied to prevent overfitting, with a patience interval of 10 epochs; i.e., the training stopped when the validation loss did not improve by more than 0.001 for 10 consecutive epochs. 
All four scenarios use the same training, validation and test set (refer to Tab.~\ref{tab:hst_ls_dist}). This allows us to draw a fair comparison of the performances for the four difference scenarios.

\section{Results and discussion}\label{section:resultsanddiscussion}

The loss curves for LS and \HST alone are shown in Fig.~\ref{fig:loss_curves_1} left panel. The training for LS required 304 epochs, while the training for \HST was completed in 327 epochs. The evolution indicate that the networks (a) based on color images from LS train more quickly than those trained on single-band image of \HST, which is expected due to the artifacts, different exposure times, and the higher resolution of \HST that encodes substantial morphological information. A summary of the performance is shown in Table~\ref{table:performance}. 

The stacked-images training loss (Fig.~\ref{fig:loss_curves_1}, right panel) shows a swift decrease, with the validation loss closely following the training loss but plateauing at a slightly higher value after around 100 epochs. This behavior suggests that the stacked-image architecture efficiently learns patterns from the training data, reaching stability relatively early but with a small generalization gap, as evidenced by the difference between the training and validation losses. The model has converged after 307 epochs, with both losses exhibiting minimal fluctuations after about 150 epochs. Contrarily, the merged-branch architecture exhibits a similar pattern, but the validation loss decreases more gradually than in the stacked-images model. The merged-branch loss continues to decrease past 150 epochs, converging after 321 epochs. This indicates that the merged architecture may require more training to fully integrate features from multiple branches, though it achieves a comparable final performance. The slower convergence could be due to the added complexity in combining independent networks. Despite this, the merged model still demonstrates competitive performance, particularly after fully integrating the features learned by the two branches.

\begin{figure}[htbt!]
    \centering
    \includegraphics[width=1.0\linewidth]{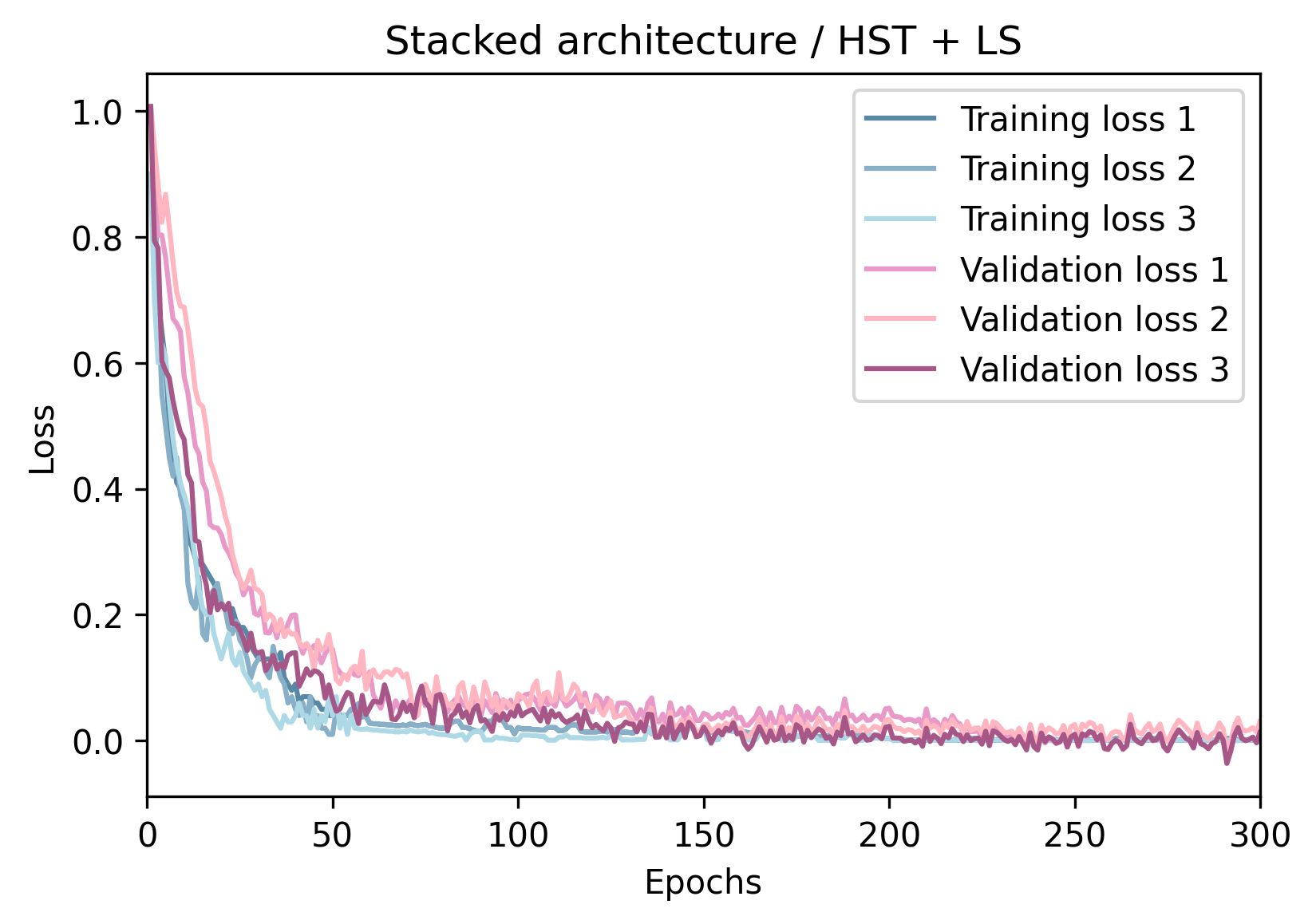}
    \caption{Learning curves for the stacked-images architecture using three different \HST and LS data sets. The data set that includes real lenses in the training (data set 3) has the best performance with the lowest loss.}
    \label{fig:test_stacked_losscurve}
\end{figure}

Since we are more interested in the stacked-images architecture, we additionally created three more learning curve tests using different training and validation data (see Fig.~\ref{fig:test_stacked_losscurve}). The first two training and their corresponding validation sets (Training loss 1, Training loss 2, Validation loss 1, and Validation loss 2) were calculated using randomly simulated data from \HST and LS in two separate training and validation data sets. Subsequently, a third evaluation (Training loss 3 and  Validation loss 3) incorporated the $\sim$360 real observational data of lens galaxy candidates identified in the literature to further refine the model's architecture. This test aimed to determine whether using only simulated data was sufficient for identifying new lenses. Importantly, the third validation set (Validation loss 3) demonstrated slightly better performance with a marginally lower loss than the first two validation sets derived from simulated data, though the improvement was not significant. Given this result, we opted to use only simulated data for training and reserved the real lens galaxy candidates for the test set.

\begin{table*}[ht!]
\centering
\caption{Summary of performance metrics for different training scenarios (see learning curves in Fig.~\ref{fig:loss_curves_1}).}
\resizebox{\textwidth}{!}{%
\begin{tabular}{|l|p{3cm}|p{3cm}|p{3cm}|p{3cm}|}
\hline
\textbf{Metric/Scenario} & \textbf{LS} & \textbf{\textit{HST}} & \textbf{Stacked images} & \textbf{Merged branches} \\
\hline
\textbf{Epochs for Training} & 304 & 327 & 307 & 321 \\
\hline
\textbf{Learning Curve Characteristics} & Fast initial learning with moderate stability post-training & Slower convergence due to noise, artifacts, and higher resolution & Swift decrease, with minimal fluctuations after ~100 epochs; stable with small generalization gap & Gradual decrease in validation loss; stable with slower convergence due to added complexity \\
\hline
\label{table:performance}
\textbf{TPR at FPR of $10^{-4}$} & 0.41 & 0.45 & 0.51 & 0.55 \\
\hline
\end{tabular}%
}
\end{table*}

\begin{figure*}
    \centering
\includegraphics[width=0.45\textwidth]{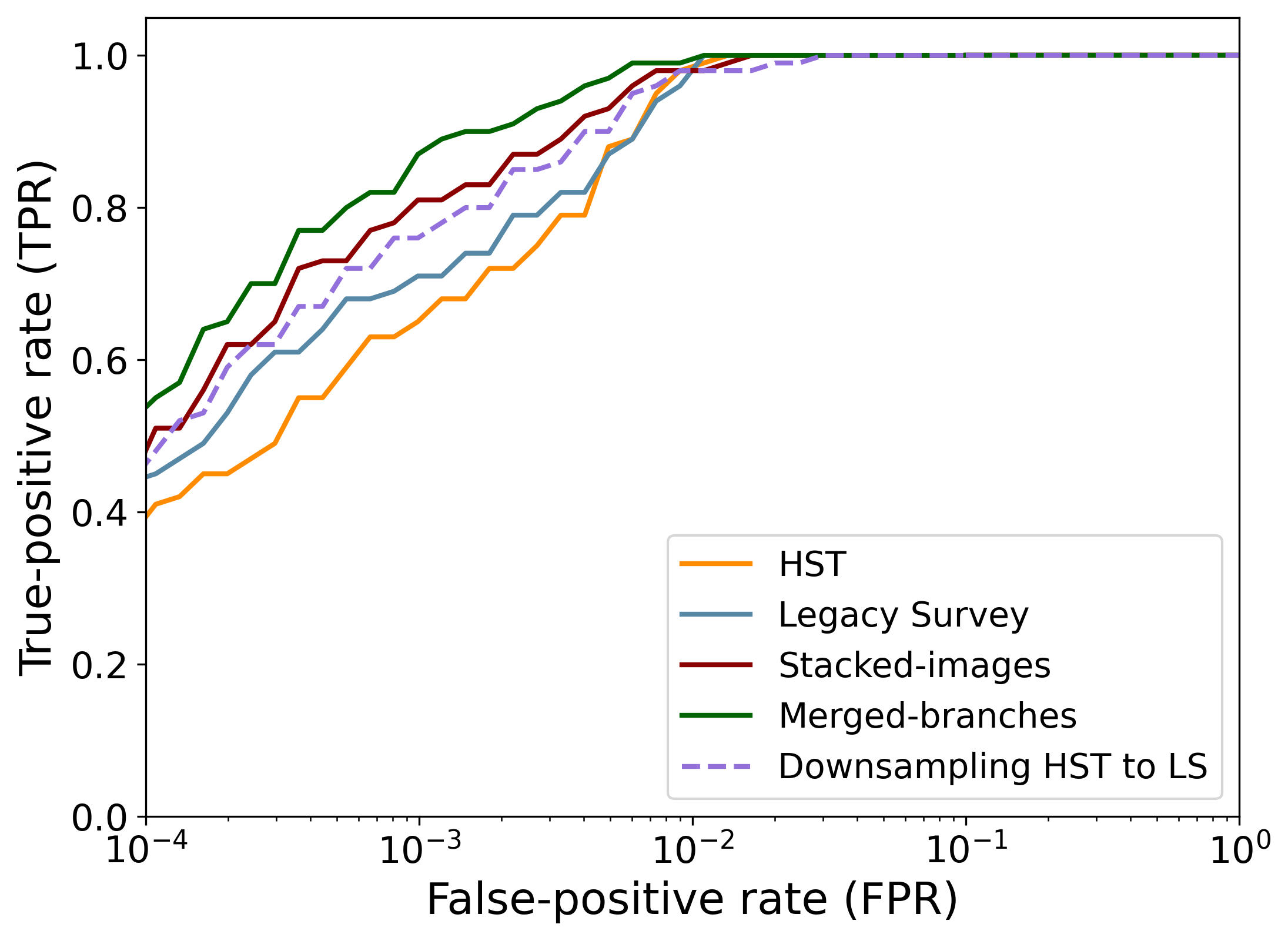}
\includegraphics[width=0.45\textwidth]{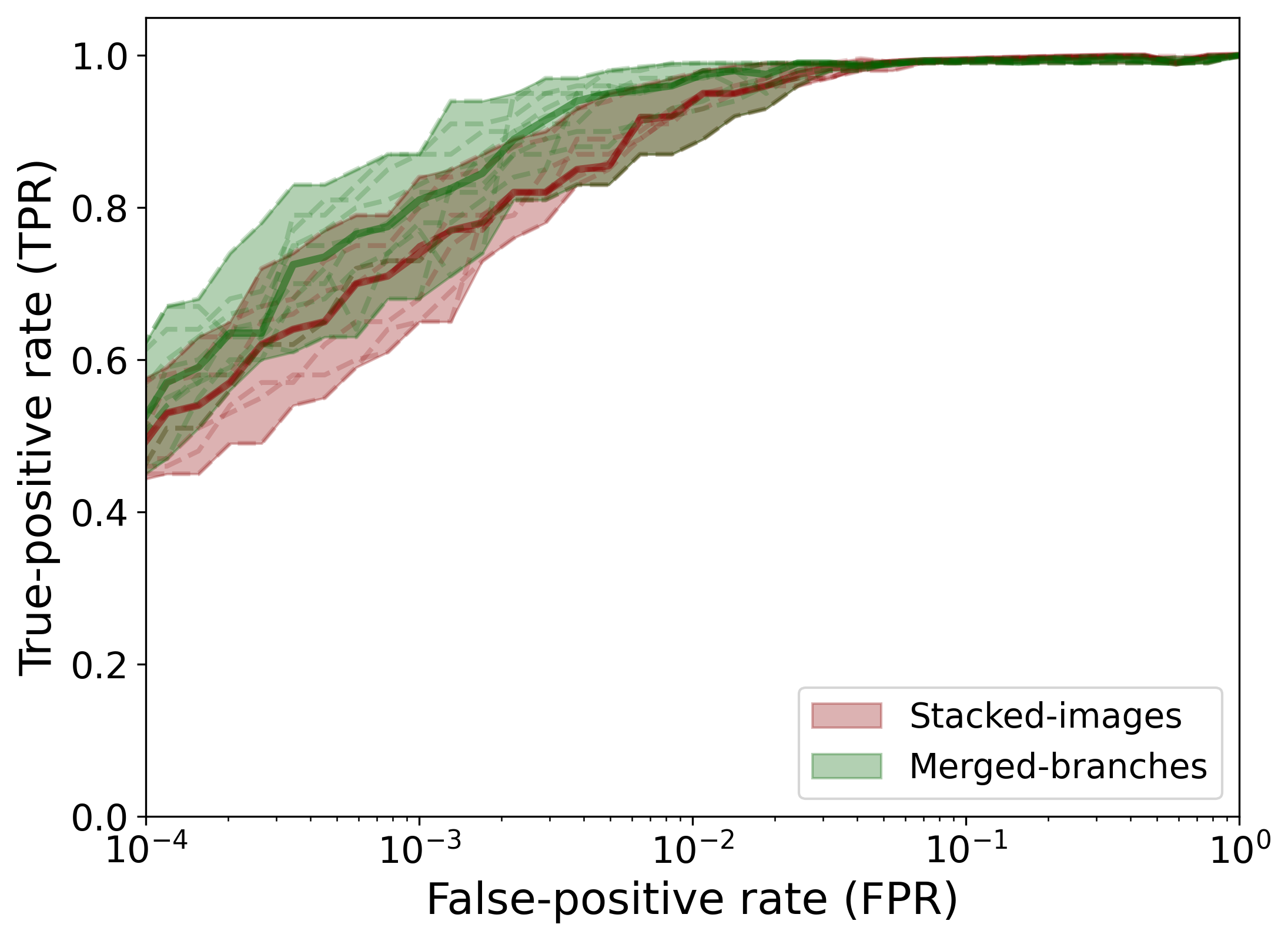}
    \caption{ROC curves of different network architectures. Left panel: our four main networks illustrated in Fig.~\ref{fig:architectures} (solid lines), and the network that downsampled \HST F814W to LS (purple dashed line). Right panel: ten different test sets for the stacked-images and merged-branches network, illustrating that the merged-branches network overall performs better than the stacked-images network.}
    \label{fig:roc_curve}
\end{figure*}

After training, we measure the performance of the four scenarios in the Receiver Operating Characteristic (ROC) curve, shown in Fig.~\ref{fig:roc_curve} left panel. This curve illustrates the relation between the true-positive rate (TPR), which indicates the model's ability to identify actual positive cases correctly, and the false-positive rate (FPR), which measures the proportion of negative cases mistakenly classified as positives. The total number of negatives is approximately 120,000 in the test set. While these sample sizes allow us to quantify low FPR values, the estimates from \HST and LS alone are less reliable in the low-FPR regime compared to the combined data set. This is primarily because HST provides high-resolution images that, while detailed, are noisier, and LS images, though offering broader wavelength coverage, suffer from lower resolution. At an FPR of $10^{-4}$, the TPR is $\sim$0.41, $\sim$0.45, $\sim$0.51 and $\sim$0.55, for \HST, LS, stacked images and merged-branches, respectively. The merged branches architecture achieves superior performance, maintaining a higher TPR at lower FPRs than the other architectures. This result underlines the potential benefits of integrating data from multiple telescopes to improve the accuracy of lens search detection. 
The performance of the \HST and LS architectures alone can be compared, for instance, to the work done by \citet{2021CañamerasHSVI} with data from the Hyper Supreme Cam (HSC), which is comparable to what LSST data will be. Their work finds a TPR of $\sim$0.60, which is higher than our \HST and LS TPRs ($\sim$0.41 and $\sim$0.45, respectively). It is important to note, however, that their test set is constructed differently, as they used HSC Wide PDR2 images, while our study uses all available real galaxy-scale lens candidates found in the literature with corresponding \HST and LS images. The image quality of HSC is expected to be comparable to that of LSST. In contrast, \HST achieves a depth of 27.2 AB magnitude in the F814W filter for a single orbit (approximately 2028 seconds of exposure). The Legacy Survey (LS), conducted using the The Dark Energy Camera Legacy Survey, reaches $5\sigma$ point source depths of approximately 24.7 AB magnitude in the g band, 23.9 AB magnitude in the r band, and 23 AB magnitude in the z band for areas with typical three-pass observations\footnote{\url{https://datalab.noirlab.edu/ls/decals}}. Accordingly, this comparison serves only as a reference point, with our primary conclusions drawn from the analysis of our four internal approaches rather than from the findings of \citet{2021CañamerasHSVI}. We attribute the higher TPR in their results to the deeper and sharper image quality of HSC in comparison to LS, and the presence of multiple bands in comparison to \HST.  

In addition, to assess the significance of the performance difference between both architectures, we have made ten different ROC curves using ten different test sets (right panel in Fig.~\ref{fig:roc_curve}). Each test set is constructed by randomly splitting the overall data set to assess the robustness and performance of the architectures. We conduct this test to determine whether the merged-branches architecture performs better than the stacked architecture or is comparable in terms of its error. Based on the analysis of ten ROC curves, we can observe a higher performance of the merged-branches architecture. On the other hand, the lower performance in the stacked-images architecture could be attributed to the difficulty in processing the four images with two different resolutions/PSFs. This aspect will be further explored in a future paper to better interpret the network's decision-making process in classifying lens and non-lens images, examining how specific features influence its predictions and the factors contributing to potential misclassifications. 

\begin{figure*}
    \centering
    \includegraphics[width=0.99\textwidth]{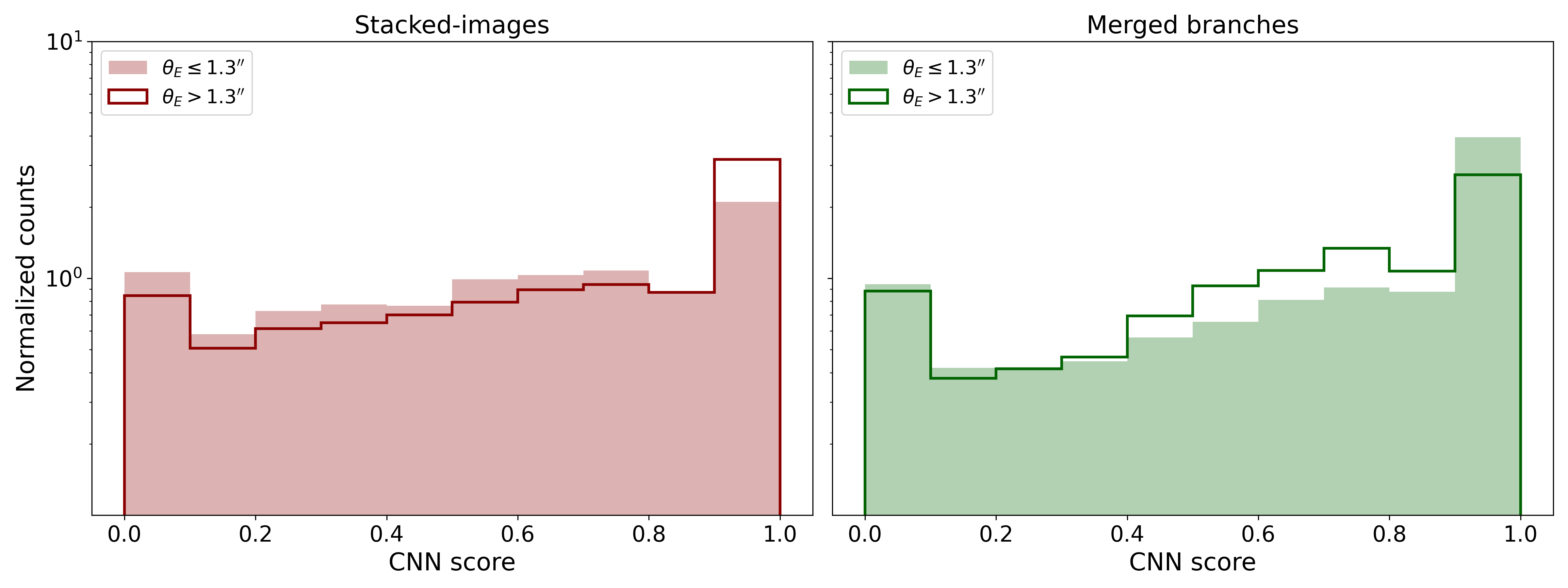}
    \caption{CNN scores for the positive examples of the test set.}
    \label{fig:ein-CNN}
\end{figure*}

To further investigate the performance differences between the architectures, we performed an additional test by downsampling \HST F814W images to match the resolution and pixel scale of the LS data set and incorporating them as a fourth LS filter. This modified LS network, using the four filters ($g$, $r$, $z$, and the downsampled F814W), was then compared to the four ROC curves mentioned previously (see Fig.~\ref{fig:roc_curve}). The results show that the performance of the "four-filter LS network" is comparable to that of the stacked-images architecture. This finding suggests that the stacked-images architecture does not fully exploit the high resolution of the \HST images. The comparable performance between the two configurations indicates that the stacked-images network may struggle to integrate the additional information provided by higher-resolution data, potentially due to complexities introduced by varying resolutions. This limitation highlights the need for further investigation into how the architecture processes and combines multi-resolution data to improve its ability to extract meaningful features.

\begin{figure}
    \centering
    \includegraphics[width=1.0\linewidth]{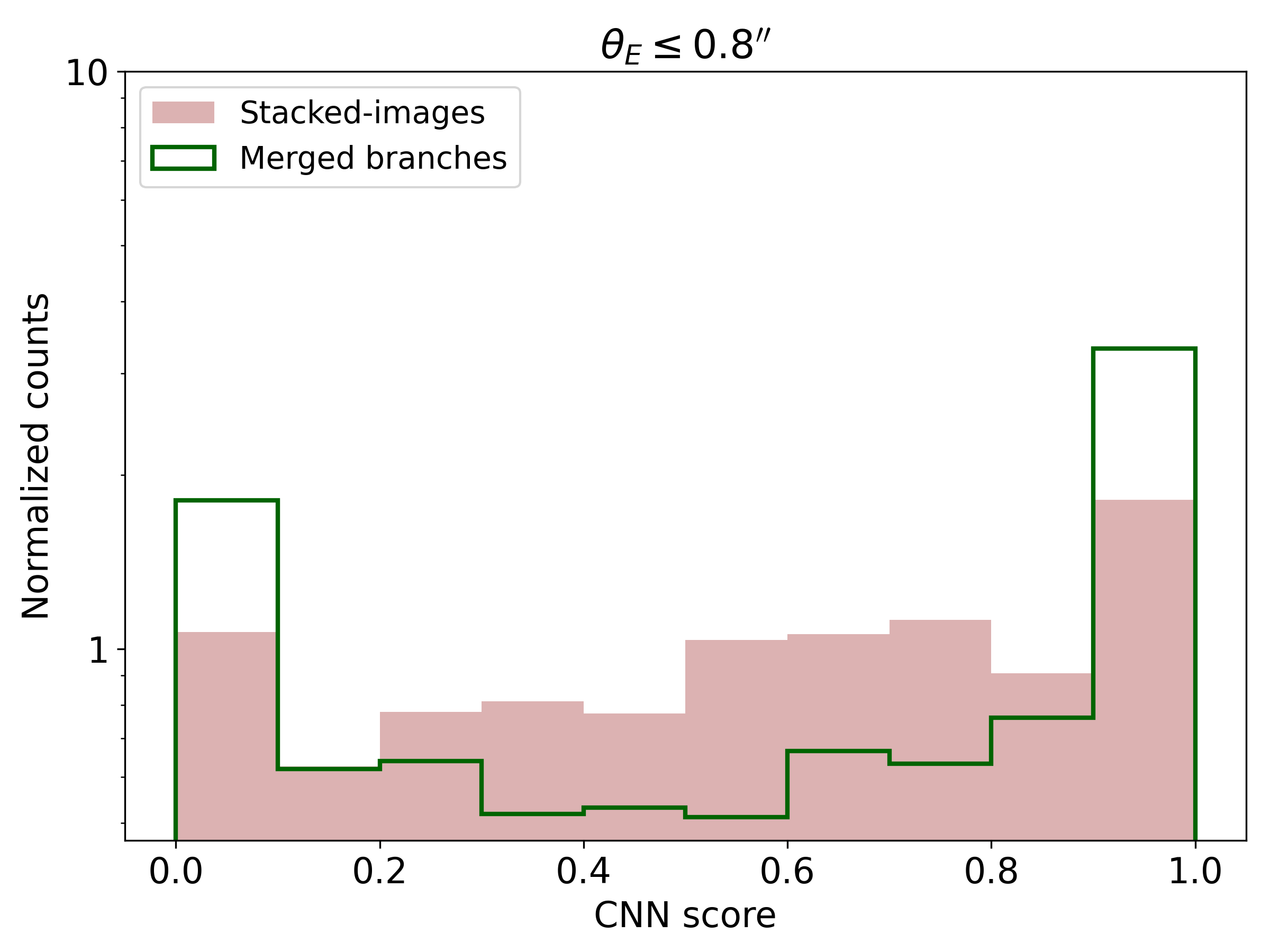}
    \caption{CNN scores for systems with $\theta_{\mathrm{E}} \leq 0.8 \arcsec$ from the test set scored by both architectures.}
    \label{fig:loweinsteinvalues}
\end{figure}

Fig.~\ref{fig:ein-CNN} compares the CNN score distributions for the positive test set between the stacked-images and merged-branches approaches. The histograms display the normalized counts of CNN scores, segmented by the $\theta_{\mathrm{E}}$ into two categories: $\theta_{\mathrm{E}} \leq$  1.3\arcsec (represented by filled histograms) and $\theta_{\mathrm{E}} >$  1.3\arcsec (represented by open histograms). Note that $65\%$ of the positive mock have $\theta_{\mathrm{E}}$ $<$ 1.3\arcsec. For the category $\theta_{\mathrm{E}} > 1.3\arcsec$, both models show a significant increase in CNN scores close to 1.0, reflecting the two architectures' ability to confidently identify these examples as positive. 

In the merged-branches model (right panel), the CNN scores for $\theta_{\mathrm{E}} \leq 1.3\arcsec$ shows a significant peak at a CNN score of 1.0. The final bin (with scores between 0.9 and 1.0) contains 2,445 mock lenses (25\% of the total positive test set), indicating that the model frequently assigns the highest confidence to these cases. Notably, 1,856 examples in this category achieved a CNN score of precisely 1.0, highlighting the model's strong tendency toward high-certainty classifications. However, for $\theta_{\mathrm{E}} > 1.3\arcsec$, it displays a less pronounced increase in CNN scores near 1.0. This suggests that, despite a subset of confident classifications, the model is generally less reliable in distinguishing lenses with larger $\theta_{\mathrm{E}}$ values.

On the other hand, the stacked-images branch (left panel of Fig.~\ref{fig:ein-CNN}) shows a lower peak at 1.0 for $\theta_{\mathrm{E}} \leq$ 1.3\arcsec, with 1,208 mock lenses in the final bin. The model shows 323 examples with a CNN score of precisely 1.0. For $\theta_{\mathrm{E}} >$ 1.3\arcsec, the stacked-images model displays 885 examples reaching a score of 1.0. However, the generally lower peaks near scores of 0.9–1 suggest that the stacked-images model may perform less confidently for lower $\theta_{\mathrm{E}}$ values compared to the merged-branches model.

Figure~\ref{fig:loweinsteinvalues} compares the performance of the stacked-images and merged-branches models on systems with low Einstein radii ($\theta_{\mathrm{E}} \leq 0.8\arcsec$). The stacked-images model shows a gradual increase in CNN scores, with counts spread across bins and peaking at 1.0. This indicates a progressive assignment of confidence, with varying confidence levels before reaching the highest score. In contrast, the merged-branches model exhibits sharper increases at specific score bins, with a strong concentration of scores at 1.0. This suggests that the merged-branches model tends to assign high-confidence scores more directly for systems it identifies as lenses. While both models peak at 1.0, the merged-branches model does so more abruptly, highlighting its preference for high-confidence classifications for systems with $\theta_{\mathrm{E}} \leq 0.8\arcsec$.

Given the total of approximately 8,000 positive examples (of which 364 are real lens candidates), the merged branches model's tendency to assign extreme confidence scores may be beneficial in scenarios prioritizing high-confidence predictions for low Einstein radius lenses. However, this focus comes at the potential cost of flexibility across a broader score distribution, as seen in the stacked-images model, which maintains a more even distribution across scores, accommodating systems with both high and low confidence levels.

\begin{figure*}
    \centering
    \includegraphics[width=0.99\textwidth]{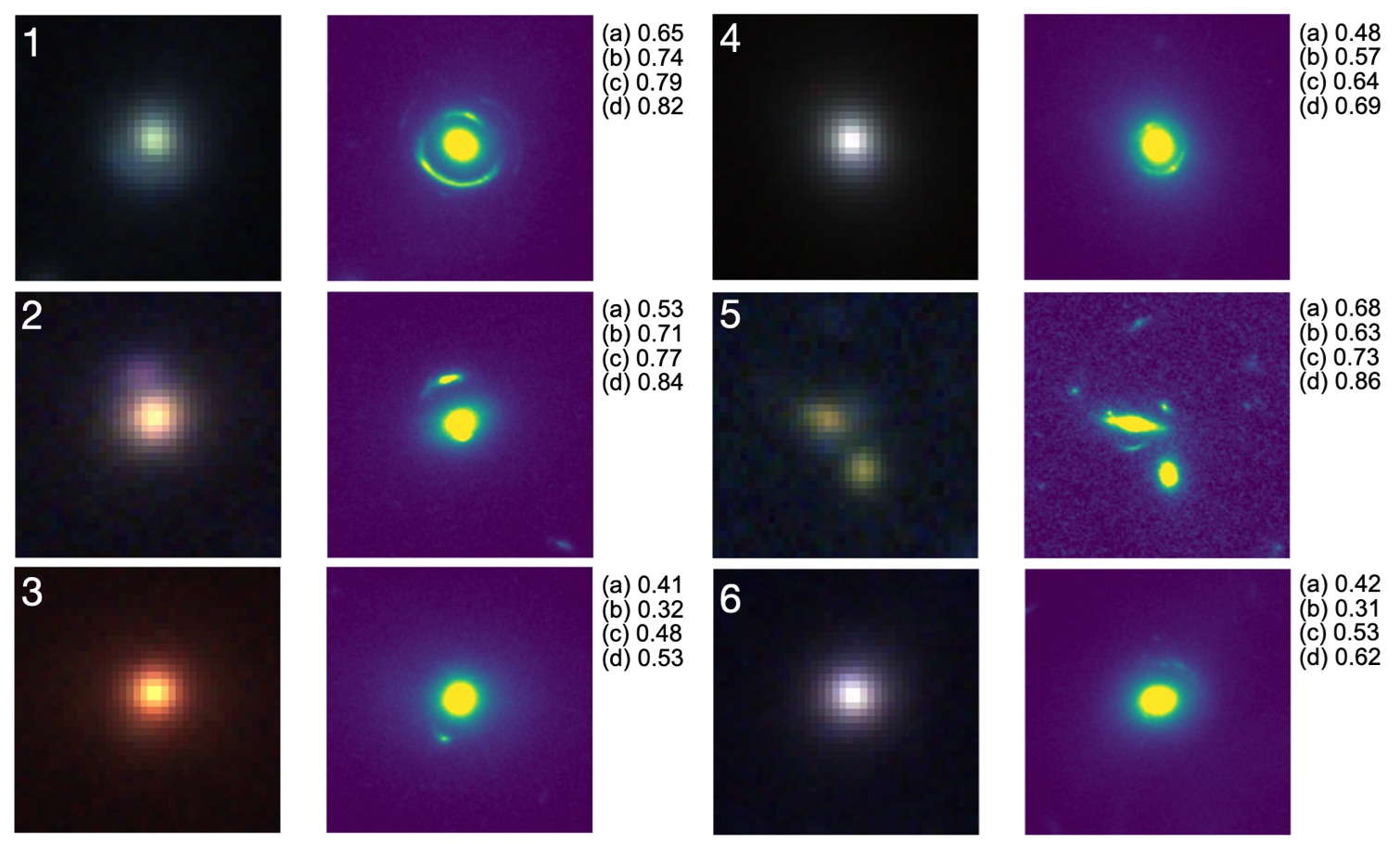}
    \caption{CNN scores of some test set examples of gravitational lensing \citep{2008Bolton,2014Pawase} for LS (RGB image) and \HST (F814W). The letters a), b), c), and d) represent the scores from the LS-only, \HST-only, stacked-images, and merged-branches network architectures.}
    \label{fig:scores}
\end{figure*}

Fig.~\ref{fig:scores} shows a selection of very likely strong lenses systems, along with the model networks' confidence in detecting lensing features: (a) LS-only, (b) \HST-only, (c) stacked-images, and (d) merged-branches. The \HST-only model generally provides higher scores compared to the the LS-only model on positive examples. This is due to the higher resolution and clearer features in the \HST images, making lensing features such as arcs and multiple images more distinguishable. However, the best performance across all cases is seen in the merged model (case d), followed closely by the stacked model (case c).

In cases where clear lensing features, such as arcs and multiple images, are visible (e.g., examples 1, 2, and 5), the scores are higher across all models, particularly in the merged and stacked models. For instance, in the 5th image, the scores for the merged model (d) and stacked model (c) reach 0.86 and 0.73, respectively, indicating a high level of confidence in detecting a lensing feature. The LS-only model (a) also performs well in this case, with a score of 0.68, but it does not reach the level of accuracy achieved by the combined models.

Conversely, for images without obvious lensing features (e.g., examples 3 and 6), even the merged and stacked models show lower confidence. This demonstrates that while combining data from different surveys improves detection, strong and clear lensing signals are still crucial for high-confidence predictions. The LS-only model (a) performs particularly poorly in these cases, with scores of 0.41 in case 3 and 0.42 in case 6, further highlighting the limitations of low-resolution data when clear features are absent.

\section{Conclusions}\label{section:Conclusions}

In this paper, we have introduced a novel architecture that combines high-resolution data from HST with multi-band, lower-resolution data from LS to improve gravitational lens detection. The first model uses only single-band, high-resolution HST images, capturing fine structural details but lacking the multi-band information available in other configurations. The second model relies solely on multi-band LS images, which provide broad coverage across three bands but at a lower resolution. The stacked
architecture combines four images ($g$, $r$, $z$ filters of LS and F814W filter of \HST) by interpolating LS images to match the pixel scale of HST and processing them simultaneously. Finally, the merged-branches architecture processes HST and LS data separately in independent branches before merging their outputs for classification.

Overall, the merged-branches architecture for \HST and LS data significantly boosts detection rates in many cases, underscoring the strength of multi-source data fusion.
Although multi-band high-resolution data would be ideal, such data sets are not yet available for large sky areas. However, our approach shows that combining high-resolution single-band data with multi-band, low-resolution images greatly enhances detection capabilities. This indicates that future lensing searches could benefit from similar multi-source data integration until multi-band high-resolution data become more widely available.

This methodology offers great potential for integrating diverse data sets, and the proposed pipeline can be applied to various types of imaging, beyond Euclid and LSST, and is not restricted to data from differing resolutions. Looking ahead, applying these networks to Euclid and LSST images could be feasible within the next few months/years, though further development and validation would be necessary to confirm this potential. For Euclid, we expect even better performance from the combination of images from different instruments, such as the high-resolution VIS images and the NISP bands, allowing for a richer data set. This approach, similar to our work of combining images of different resolution, should further improve the detection capabilities compared to \HST and LS alone. Additionally, LSST’s multiband data will provide complementary coverage, further enhancing performance. 

\begin{acknowledgements}
SHS thanks the Max Planck Society for support through the Max Planck Fellowship. This project has received funding from the European Research Council (ERC) under the European Union's Horizon 2020 research and innovation programme (grant agreement No 771776).  This research is supported in part by the Excellence Cluster ORIGINS which is funded by the Deutsche Forschungsgemeinschaft (DFG, German Research Foundation) under Germany's Excellence Strategy -- EXC-2094 -- 390783311. SS has received funding from the European Union’s Horizon 2022 research and innovation programme under the Marie Skłodowska-Curie grant agreement No
101105167 — FASTIDIoUS. SB acknowledges the funding provided by the Alexander von Humboldt Foundation.
\end{acknowledgements}

%
%

\bibliographystyle{aa} 
\bibliography{references}

\end{document}